\documentclass[journal]{IEEEtran}
\usepackage[T1]{fontenc}
\usepackage{ifpdf}
\usepackage{cite}
\usepackage[pdftex]{graphicx}
\usepackage{amsmath}
\usepackage{amssymb}
\usepackage{bm}
\usepackage{array}
\usepackage{fixltx2e}
\usepackage{stfloats}

\usepackage{diagbox}

\usepackage{makecell}

\usepackage{threeparttable}

\usepackage{color,soul}
\soulregister\cite7

\usepackage[hidelinks]{hyperref}

\begin{document}
\title{Globally Optimal Selection of Ground Stations in Satellite Systems with Site Diversity}

\author{Christos~N.~Efrem, and~Athanasios~D.~Panagopoulos,~\IEEEmembership{Senior~Member,~IEEE}
\thanks{C. N. Efrem and A. D. Panagopoulos are with the School of Electrical and Computer Engineering, National Technical University of Athens, 15780 Athens, Greece (e-mails: chefr@central.ntua.gr, thpanag@ece.ntua.gr).

This article has been accepted for publication in \textit{IEEE Wireless Communications Letters}, DOI: 10.1109/LWC.2020.2982139.  Copyright \textcopyright \ 2020 IEEE. Personal use is permitted, but republication/redistribution requires IEEE permission. See \url{http://www.ieee.org/publications_standards/publications/rights/index.html} for more information.
}}

\markboth{}%
{}

\maketitle

\begin{abstract}
The availability of satellite communication systems is extremely limited by atmospheric impairments, such as rain (for radio frequencies) and cloud coverage (for optical frequencies). A solution to this problem is the site diversity technique, where a network of geographically distributed ground stations (GSs) can ensure, with high probability, that at least one GS is available for connection to the satellite at each time period. However, the installation of redundant GSs induces unnecessary additional costs for the network operator. In this context, we study an optimization problem that minimizes the number of required GSs, subject to availability constraints. First, the problem is transformed into a binary-integer-linear-programming (BILP) problem, which is proven to be NP-hard. Subsequently, we design a branch-and-bound (B\&B) algorithm, with global-optimization guarantee, based on the linear-programming (LP) relaxation and a greedy method as well. Finally, numerical results show that the proposed algorithm significantly outperforms state-of-the-art methods, and has low complexity in the average case.

\end{abstract}


\IEEEpeerreviewmaketitle

\section{Introduction}
\IEEEPARstart{S}{ite} diversity technique is used to improve the availability of satellite communication systems by mitigating the atmospheric effects \cite{Panagopoulos_FMT}. In particular, multiple GSs separated over long distances receive the same signal from the satellite, and in this way the probability of all GSs experiencing severe weather conditions simultaneously is reduced. A joint optimization method for the design of optical satellite networks is proposed in \cite{Poulenard}, which consists of two parts. The first part is the optical-GS positioning optimization performed by an iterative greedy procedure, while the second part is the backbone network optimization taking into consideration the optical fiber cost. In \cite{Fuchs}, a network optimization method with reduced complexity is presented, exploiting the single-site availabilities as well as the correlation between sites. 

Furthermore, the optimal location of optical GSs for low-earth-orbit (LEO) satellite missions is examined in \cite{Portillo} through multi-objective optimization, using genetic algorithms (GAs) and considering three performance metrics: system availability, latency, and network cost. GAs are also used in \cite{Rossi_optimization} to minimize two different objective functions in extremely-high-frequency (EHF) satellite networks with smart-gateway (SG) diversity. In addition, the selection of the minimum number of GSs in optical satellite networks with a medium-earth-orbit (MEO) or a geostationary (GEO) satellite is investigated in \cite{Lyras_MEO} and \cite{Lyras_GEO}, respectively. Both studies present heuristic algorithms of low complexity, taking into account the spatial correlation as well as the monthly variability of cloud coverage.

The \textit{main contributions} of this work compared to existing approaches are the following: 1) rigorous mathematical \linebreak formulation of the optimization problem with a formal proof of its NP-hardness, 2) system availability guarantee for several time periods (e.g., months), not only for a year, and 3) unlike existing methods that provide suboptimal solutions without any performance guarantee, the designed B\&B algorithm achieves global optimality with low average-case complexity (i.e., good trade-off between performance and complexity).

The remainder of this article is structured as follows. Section II presents the system model and formulates the optimization problem, which is transformed into an NP-hard BILP problem in Section III. Afterwards, a global optimization algorithm is given in Section IV, while its performance is numerically analyzed in Section V. Finally, Section VI concludes the paper.

\textit{Mathematical notation}: The absolute value of a real number $x$ is denoted by $\left| x \right|$, while $\left| \mathcal{D} \right| = D$ represents the cardinality of a set $\mathcal{D}$. Also, ${{\mathbf{0}}_N}$/${{\mathbf{1}}_N}$ stands for the $N$-dimensional zero/\linebreak all-ones vector respectively, and $\left\lceil  \cdot  \right\rceil$ is the ceiling function. 

\section{System Model and Problem Formulation}
We consider a satellite system with a geostationary satellite and a ground station network employing site diversity. Specifically, $\mathcal{K} = \{ 1,2, \ldots ,K\}$ is the set of available locations for installing a GS (or, equivalently, the set of candidate GSs), and $\mathcal{T} = \{ 1,2, \ldots ,T\}$ denotes the set of time periods (e.g., months). In addition, $p_{k,t}^{{\text{out}}}$ is the outage probability of GS $k$ in time period $t$,\footnote{In \textit{radio-frequency (RF) satellite systems}, a GS is in outage when the rain attenuation exceeds a specific threshold \cite{Rossi_optimization}, which is determined by the required bit-error-rate (BER). In \textit{optical satellite networks}, a GS is in outage when experiencing cloud blockage \cite{Sanchez, Lyras_CFLOS}. Otherwise, the GS is available.} and $P_t^{{\text{out,req}}}$ is the maximum required system outage probability in time period \nolinebreak $t$. 

Moreover, we make the following \textit{assumptions}: a) $\{p_{k,t}^{{\text{out}}}\}_{k \in \mathcal{K}}$ are probabilities of \textit{mutually independent} events, $\forall t \in \mathcal{T}$,\footnote{This can be achieved if the distance between any two distinct GSs is sufficiently large, and therefore the spatial correlation of weather conditions is negligible. Furthermore, this case is quite common and preferable in practice, so as to take full advantage of site diversity by attaining the highest availability.} b) the system availability is defined as \textit{the probability of having at least one GS available}, c) $\{p_{k,t}^{{\text{out}}}\}_{ k \in \mathcal{K}, \, t \in \mathcal{T} }$ are supposed to be accurate (i.e., \textit{without uncertainty}); the uncertainty in the calculation of outage probabilities is beyond the scope of this paper, and d) without loss of generality we assume that $p_{k,t}^{{\text{out}}},P_t^{{\text{out,req}}} > 0$, $\forall k \in \mathcal{K}$ and $\forall t \in \mathcal{T}$.

In order to reduce the cost of installing and operating the GSs, we study the following \textit{cardinality minimization problem}:
\begin{equation} \label{initial_problem}
\begin{aligned}
  & \mathop {\min }\limits_{\mathcal{S} \subseteq \mathcal{K}} && \left| \mathcal{S} \right| = S  \\
  & \,\,\text{s.t.} && \,P_t^{{\text{avl}}}(\mathcal{S}) \geq P_t^{{\text{avl,req}}},\;\;\forall t \in \mathcal{T}  
\end{aligned}
\end{equation}
where $\mathcal{S}$ denotes the set of selected GSs, $P_t^{{\text{avl}}}(\mathcal{S}) = 1 - \prod\limits_{s \in \mathcal{S}} {p_{s,t}^{{\text{out}}}}$ is the system availability in time period $t$ achieved by the set $\mathcal{S}$ of GSs (or, equivalently, the probability of having at least one GS in $\mathcal{S}$ available in time period $t$), and $P_t^{{\text{avl,req}}} = 1 - P_t^{{\text{out,req}}}$ is the minimum required system availability in time period $t$. Notice that      $P_t^{{\text{avl}}}(\mathcal{S}) \geq P_t^{{\text{avl,req}}}$ $\Leftrightarrow$ $\prod\limits_{s \in \mathcal{S}} {p_{s,t}^{{\text{out}}}} \leq P_t^{{\text{out,req}}}$, $\forall t \in \mathcal{T}$.

\section{Equivalent BILP Problem and NP-hardness}
Subsequently, we introduce the vector ${\mathbf{z}} = [{z_1},{z_2}, \ldots ,{z_K}]$ of binary (0-1) variables. In particular, ${z_k} = 1$ if $k \in \mathcal{S}$, i.e., the ${k^{{\text{th}}}}$ GS is selected (or, equivalently, a GS is installed at the ${k^{{\text{th}}}}$ location), otherwise ${z_k} = 0$. Based on this definition, we have that $\left| \mathcal{S} \right| = \sum\limits_{k \in \mathcal{K}} {{z_k}}$ and $\prod\limits_{s \in \mathcal{S}} {p_{s,t}^{{\text{out}}}}  = {\prod\limits_{k \in \mathcal{K}} {( {p_{k,t}^{{\text{out}}}} )} ^{{z_k}}}$. As a result, problem \eqref{initial_problem} can be written as follows: 
\begin{equation}
\begin{aligned}
  & \mathop {\min }\limits_{\mathbf{z}} && \sum\limits_{k \in \mathcal{K}} {{z_k}}  \\
  & \,\,\text{s.t.} && \,{\prod\limits_{k \in \mathcal{K}} {( {p_{k,t}^{{\text{out}}}} )} ^{{z_k}}} \leq P_t^{{\text{out,req}}},\;\;\forall t \in \mathcal{T} \\
  &&& \,{z_k} \in \{ 0,1\} ,\;\;\forall k \in \mathcal{K}
\end{aligned}
\end{equation}

By taking the logarithms on both sides of the inequality-constraints and then multiplying by $-1$, we obtain an equivalent BILP problem:
\begin{equation} \label{original_BILP_problem}
\begin{aligned}
  & \mathop {\min }\limits_{\mathbf{z}} && g({\mathbf{z}}) = \sum\limits_{k \in \mathcal{K}} {{z_k}}  \\
  & \,\,\text{s.t.} && \!\sum\limits_{k \in \mathcal{K}} {{\alpha _{t,k}}{z_k}}  \geq {\beta _t},\;\;\forall t \in \mathcal{T} \\
  &&& {z_k} \in \{ 0,1\} ,\;\;\forall k \in \mathcal{K}
\end{aligned}
\end{equation}
with ${\alpha _{t,k}} = \log ( 1/{p_{k,t}^{{\text{out}}}} )$ and ${\beta _t} = \log ( 1/{P_t^{{\text{out,req}}}} )$, $\forall t \in \mathcal{T}$ and $\forall k \in \mathcal{K}$. Note that ${\alpha _{t,k}},{\beta _t} \geq 0$, since $0 < p_{k,t}^{{\text{out}}},P_t^{{\text{out,req}}} \leq 1$. 

\vspace{1mm}
\newtheorem{theorem}{Theorem}
\begin{theorem}
The equivalent BILP problem \eqref{original_BILP_problem} is NP-hard.
\end{theorem}
\vspace{1mm}

\renewcommand{\IEEEQED}{\IEEEQEDopen}
\begin{IEEEproof}
In order to prove the NP-hardness of problem \eqref{original_BILP_problem}, the following property is exploited: if a special case of a problem is NP-hard, so is the general problem. Now, we consider the \textit{minimum node cover problem} (MNCP): Given a graph $G({\mathcal{N},\mathcal{E}})$, with $\mathcal{N}$ and $\mathcal{E}$ being the sets of nodes and edges respectively, find a minimum-cardinality set of nodes $\mathcal{N}' \subseteq \mathcal{N}$ such that $\{ n,m\} \in \mathcal{E}$ $\Rightarrow$ $n \in \mathcal{N}'$ or $m \in \mathcal{N}'$. Furthermore, the MNCP is known to be NP-hard \cite{Papadimitriou} and can be formulated as the following BILP problem: 
\begin{equation} \label{MNCP}
\begin{aligned}
  & \mathop {\min }\limits_{\mathbf{z}} && \sum\limits_{n \in \mathcal{N}} {{z_n}}  \\
  & \,\,\text{s.t.} && \, {z_n} + {z_m} \geq 1,\;\;\forall \{ n,m\}  \in \mathcal{E} \\
  &&& \, {z_n} \in \{ 0,1\} ,\;\;\forall n \in \mathcal{N}
\end{aligned}
\end{equation}
Obviously, the NP-hard problem \eqref{MNCP} constitutes a special case of the general problem \eqref{original_BILP_problem}, and so we have Theorem 1.
\end{IEEEproof}

\section{Global Optimization Algorithm}
Since problem \eqref{original_BILP_problem} is proven to be NP-hard, it cannot be solved in polynomial time unless P$=$NP. In other words, it is rather unlikely that there is an algorithm which finds an optimal solution and has polynomial complexity in the worst case. Nevertheless, we will design a global optimization B\&B algorithm of low average-case complexity. B\&B is an intelligent technique which recursively splits the search space into smaller spaces (\textit{branching}), and uses appropriate bounds on the optimum value (\textit{bounding}) to avoid, as much as possible, the exhaustive enumeration of candidate solutions \nolinebreak \cite{Papadimitriou}.

Next, consider problem \eqref{original_BILP_problem} with some variables being fixed:
\begin{equation} \label{BILP_subproblem}
\begin{aligned}
  & \mathop {\min }\limits_{{{\mathbf{z}}_\mathcal{V}}} && g({{\mathbf{z}}_\mathcal{V}};{{{\mathbf{\bar z}}}_\mathcal{C}}) = \sum\limits_{v \in \mathcal{V}} {{z_v}}  + \sum\limits_{c \in \mathcal{C}} {{{\bar z}_c}}  \\
  & \,\,\text{s.t.} && \!\sum\limits_{v \in \mathcal{V}} {{\alpha _{t,v}}{z_v}}  \geq {\beta'_t},\;\;\forall t \in \mathcal{T} \\
  &&& {z_v} \in \{ 0,1\} ,\;\;\forall v \in \mathcal{V}
\end{aligned}
\end{equation}
where the sets $\mathcal{V}$ and $\mathcal{C}$ contain the indices of free and constant variables respectively ($\mathcal{V} \cup \mathcal{C} = \mathcal{K}$, $\mathcal{V} \cap \mathcal{C} = \emptyset$), ${{\mathbf{z}}_\mathcal{V}} = {[{z_v}]_{v \in \mathcal{V}}}$, ${{\mathbf{\bar z}}_\mathcal{C}} = {[{\bar z_c}]_{c \in \mathcal{C}}}$ (with ${\bar z_c} \in \{ 0,1\}$, $\forall c \in \mathcal{C}$), and ${\beta'_t} = {\beta _t} - \sum\limits_{c \in \mathcal{C}} {{\alpha _{t,c}}{{\bar z}_c}}$, $\forall t \in \mathcal{T}$. Notice that when $\mathcal{V} = \mathcal{K}$ and $\mathcal{C} = \emptyset$, problem \eqref{BILP_subproblem} is identical to the original problem \eqref{original_BILP_problem}. Also, ${\mathbf{z}}_\mathcal{V}^{{\text{opt}}}$ denotes an optimal solution of problem \eqref{BILP_subproblem}, and $\left| \mathcal{V} \right| = V \leq K$.

Moreover, the following statements can be easily proven: a) $g^* \leq g({\mathbf{z}}_\mathcal{V}^{{\text{opt}}};{{\mathbf{\bar z}}_\mathcal{C}})$, where $g^*$ is the optimum value of problem \nolinebreak \eqref{original_BILP_problem}, b) if ${{\mathbf{z}}_\mathcal{V}}$ is a feasible solution of problem \eqref{BILP_subproblem}, then $[{{\mathbf{z}}_\mathcal{V}};{{\mathbf{\bar z}}_\mathcal{C}}]$ is a feasible solution of problem \eqref{original_BILP_problem}, and c) \textit{necessary-and-sufficient feasibility condition}: problem \eqref{BILP_subproblem} is feasible $\Leftrightarrow$ $\sum\limits_{v \in \mathcal{V}} {{\alpha _{t,v}}}  \geq {\beta'_t}$, $\forall t \in \mathcal{T}$ (i.e., ${{\mathbf{1}}_V}$ is a feasible solution). 

Now, in order to construct a \textit{lower bound} on the optimum value of problem \eqref{BILP_subproblem}, the \textit{LP relaxation} is exploited, where the binary constraints (${z_v} \in \{ 0,1\}$, $\forall v \in \mathcal{V}$) are relaxed:
\begin{equation} \label{LP_relaxation}
\begin{aligned}
  & \mathop {\min }\limits_{{{\mathbf{z}}_\mathcal{V}}} && g({{\mathbf{z}}_\mathcal{V}};{{{\mathbf{\bar z}}}_\mathcal{C}}) = \sum\limits_{v \in \mathcal{V}} {{z_v}}  + \sum\limits_{c \in \mathcal{C}} {{{\bar z}_c}}  \\
  & \,\,\text{s.t.} && \!\sum\limits_{v \in \mathcal{V}} {{\alpha _{t,v}}{z_v}}  \geq {\beta'_t},\;\;\forall t \in \mathcal{T} \\
  &&& 0 \leq {z_v} \leq 1,\;\;\forall v \in \mathcal{V}
\end{aligned}
\end{equation}
An optimal solution ${\mathbf{z}}_\mathcal{V}^{{\text{LP}}}$ of the LP relaxation (problem \eqref{LP_relaxation}) can be obtained in polynomial time, using interior-point methods \cite{Ben-Tal}. In addition, note that: a) the feasibility of problem \eqref{BILP_subproblem} implies the feasibility of the LP relaxation, \linebreak b) if ${\mathbf{z}}_\mathcal{V}^{{\text{LP}}} \in {\{ 0,1\} ^V}$, then $g({\mathbf{z}}_\mathcal{V}^{{\text{opt}}};{{\mathbf{\bar z}}_\mathcal{C}}) = g({\mathbf{z}}_\mathcal{V}^{{\text{LP}}};{{\mathbf{\bar z}}_\mathcal{C}})$, and c) $g({\mathbf{z}}_\mathcal{V}^{{\text{LP}}};{{\mathbf{\bar z}}_\mathcal{C}}) \leq g({\mathbf{z}}_\mathcal{V}^{{\text{opt}}};{{\mathbf{\bar z}}_\mathcal{C}})$, and because $g({\mathbf{z}}_\mathcal{V}^{{\text{opt}}};{{\mathbf{\bar z}}_\mathcal{C}})$ is an integer, we have that $\left\lceil {g({\mathbf{z}}_\mathcal{V}^{{\text{LP}}};{{{\mathbf{\bar z}}}_\mathcal{C}})} \right\rceil  \leq g({\mathbf{z}}_\mathcal{V}^{{\text{opt}}};{{\mathbf{\bar z}}_\mathcal{C}})$. 

In the sequel, we develop a greedy method to provide an \textit{upper bound} on the optimum value of problem \eqref{BILP_subproblem}. This method is based on the following \textit{cost function} (CF): $f({{\mathbf{z}}_\mathcal{V}}) = \sum\limits_{t \in \mathcal{T}} {\max ({{d_t},0})}$, with ${d_t} = {\beta '_t} - \sum\limits_{v \in \mathcal{V}} {{\alpha _{t,v}}{z_v}}$, $\forall t \in \mathcal{T}$, which quantifies the total violation of inequality-constraints induced by the vector ${{\mathbf{z}}_\mathcal{V}}$. Observe that: a) $f({{\mathbf{z}}_\mathcal{V}}) \geq 0$, and b) $f({{\mathbf{z}}_\mathcal{V}}) = 0$ $\Leftrightarrow$ $\sum\limits_{v \in \mathcal{V}} {{\alpha _{t,v}}{z_v}}  \geq {\beta '_t}$, $\forall t \in \mathcal{T}$.  

\begin{table}[!t]
\centering
\renewcommand{\arraystretch}{1.35}
\begin{tabular*}{\columnwidth}{@{}l@{}}
\hline
\textbf{\normalsize{Algorithm 1}} \normalsize{CF-based greedy method}
\\ \hline
\textbf{Input:} The BILP problem \eqref{BILP_subproblem} with $\sum\limits_{v \in \mathcal{V}} {{\alpha _{t,v}}}  \geq {\beta '_t}$, $\forall t \in \mathcal{T}$ \\
\textbf{Output:} A feasible solution ${{\mathbf{z}}_\mathcal{V}}$ of problem \eqref{BILP_subproblem} \\
\begin{tabular}{@{}r@{~}l@{}}
1: & ${{\mathbf{z}}_\mathcal{V}}: = {{\mathbf{0}}_V}$, $\mathcal{R}: = \mathcal{V}$, ${d_t} = {\beta '_t} \;\; \forall t \in \mathcal{T}$, $f: = \sum\limits_{t \in \mathcal{T}} {\max ({{d_t},0})}$ \\
2: & \textbf{while} $f>0$ \textbf{do}  \\
3: & ~~$n: = \mathop {\arg \min }\limits_{r \in \mathcal{R}} \sum\limits_{t \in \mathcal{T}} {\max ({{d_t} - {\alpha _{t,r}},0})}$, ${z_n}: = 1$, $\mathcal{R}: = \mathcal{R}\backslash \{ n\}$ \\                                                                                                                                            
4: & ~~${d_t}: = {d_t} - {\alpha _{t,n}} \;\; \forall t \in \mathcal{T}$, $f: = \sum\limits_{t \in \mathcal{T}} {\max ({{d_t},0})}$ \\
5: & \textbf{end while}  \\       
\end{tabular}
\\ \hline
\end{tabular*}
\vspace{-4mm}
\end{table}

Algorithm 1 presents the CF-based greedy method, where $\mathcal{R} = \{ v \in \mathcal{V}:\,{z_v} = 0\}$. In particular, ${{\mathbf{z}}_\mathcal{V}}$ is initialized to the zero vector, and in each iteration we find the index in $\mathcal{R}$ which minimizes the CF when the corresponding 0-variable changes to 1. Then, this variable is set equal to 1 and its index is removed from the set $\mathcal{R}$. The algorithm terminates when the CF equals 0, i.e., all the inequality-constraints are satisfied. In addition, $g({\mathbf{z}}_\mathcal{V}^{{\text{opt}}};{{\mathbf{\bar z}}_\mathcal{C}}) \leq g({\mathbf{z}}_\mathcal{V}^{{\text{CF}}};{{\mathbf{\bar z}}_\mathcal{C}})$, where ${\mathbf{z}}_\mathcal{V}^{{\text{CF}}}$ is a feasible solution of problem \eqref{BILP_subproblem} obtained from Algorithm 1.

\textit{Complexity of Algorithm 1}: The complexity of the ${i^{{\text{th}}}}$ iteration is $\Theta({T(V + 1 - i)})$, since $\left| \mathcal{R} \right| = V + 1 - i$. From the input assumption of Algorithm 1 (feasibility condition), we have that $f({{\mathbf{1}}_V}) = 0$, and therefore Algorithm 1 requires a maximum of $V$ iterations to terminate. Consequently, the worst-case complexity of Algorithm 1 is $\sum\limits_{i = 1}^V {T(V + 1 - i)}  = T\sum\limits_{j = 1}^V j  = TV(V + 1)/2 = \Theta( {T{V^2}} )$, i.e., polynomial in the size of the input. 

As concerns the branching procedure in the B\&B method, we choose a \textit{branching variable} ${z_b}$ ($b \in \mathcal{V}$) such that $z_b^{{\text{LP}}}$ is the most ``uncertain'' fractional variable, i.e., closer to 0.5 than any other variable in ${\mathbf{z}}_\mathcal{V}^{{\text{LP}}}$. Afterwards, problem \eqref{BILP_subproblem} is decomposed into two \textit{subproblems} by setting either ${z_b} = 0$ or ${z_b} = 1$:
\begin{equation} \label{BILP_subproblem_0}
\begin{aligned}
  & \mathop {\min }\limits_{{{\mathbf{z}}_{\mathcal{V}\backslash \{ b\} }}} && g({{\mathbf{z}}_{\mathcal{V}\backslash \{ b\} }};{{{\mathbf{\bar z}}}_{\mathcal{C} \cup \{ b\} }}) = \sum\limits_{v \in \mathcal{V}\backslash \{ b\} } {{z_v}}  + \sum\limits_{c \in \mathcal{C}} {{{\bar z}_c}}  \\
  & \,\,\,\,\text{s.t.} && \!\sum\limits_{v \in \mathcal{V}\backslash \{ b\} } {{\alpha _{t,v}}{z_v}}  \geq {\beta'_t},\;\;\forall t \in \mathcal{T} \\
  &&& {z_v} \in \{ 0,1\} ,\;\;\forall v \in \mathcal{V}\backslash \{ b\}
\end{aligned}
\end{equation}

\begin{equation} \label{BILP_subproblem_1}
\begin{aligned}
  & \mathop {\min }\limits_{{{\mathbf{z}}_{\mathcal{V}\backslash \{ b\} }}} && g({{\mathbf{z}}_{\mathcal{V}\backslash \{ b\} }};{{{\mathbf{\bar z}}}_{\mathcal{C} \cup \{ b\} }}) = \sum\limits_{v \in \mathcal{V}\backslash \{ b\} } {{z_v}}  + \sum\limits_{c \in \mathcal{C}} {{{\bar z}_c}} + 1  \\
  & \,\,\,\,\text{s.t.} && \!\sum\limits_{v \in \mathcal{V}\backslash \{ b\} } {{\alpha _{t,v}}{z_v}}  \geq {\beta'_t} - {\alpha _{t,b}},\;\;\forall t \in \mathcal{T} \\
  &&& {z_v} \in \{ 0,1\} ,\;\;\forall v \in \mathcal{V}\backslash \{ b\}
\end{aligned}
\end{equation}
These subproblems have the same form as problem \eqref{BILP_subproblem}, with ${\bar z_b} = 0$/$1$ for subproblem \eqref{BILP_subproblem_0}/\eqref{BILP_subproblem_1}, respectively. Moreover, if $g_0^\text{opt}$ and $g_1^\text{opt}$ are respectively the optimum values of subproblems \eqref{BILP_subproblem_0} and \eqref{BILP_subproblem_1} (assuming that the optimum value of an infeasible problem equals $+ \infty$), then $g({\mathbf{z}}_\mathcal{V}^{{\text{opt}}};{{\mathbf{\bar z}}_\mathcal{C}}) = \min (g_0^\text{opt} ,g_1^\text{opt})$.

\begin{table}[!t]
\centering
\renewcommand{\arraystretch}{1.35}
\begin{tabular*}{\columnwidth}{@{}l@{}}
\hline
\textbf{\normalsize{Algorithm 2}} \normalsize{LP\&CF-based B\&B method }
\\ \hline
\textbf{Input:} The original BILP problem \eqref{original_BILP_problem} with $\sum\limits_{k \in \mathcal{K}} {{\alpha _{t,k}}}  \geq {\beta _t}$, $\forall t \in \mathcal{T}$ \\
\textbf{Output:} A (globally) optimal solution ${{\mathbf{z}}^*}$ of problem \eqref{original_BILP_problem} \\
\begin{tabular}{@{}r@{~}l@{}}
1: & ${{\mathbf{z}}^*}: = {{\mathbf{1}}_K}$, $U: = K$, $\mathcal{L}: = \{ \text{problem \eqref{original_BILP_problem}} \}$ \\
2: & \textbf{while} $\mathcal{L} \ne \emptyset$ \textbf{do}  \\
3: & ~~Remove the front subproblem from the list $\mathcal{L}$,  \\
   & ~~which has the form of problem \eqref{BILP_subproblem} \\ 
4: & ~~\textbf{if} $\exists t \in \mathcal{T}: \sum\limits_{v \in \mathcal{V}} {{\alpha _{t,v}}} < {\beta'_t}$ \textbf{then} \{\textbf{continue}\} \textbf{end if}    \hfill $\triangleright$ \textit{Infeasibility} \\
5: & ~~Compute an optimal solution ${\mathbf{z}}_\mathcal{V}^{{\text{LP}}}$ of the LP relaxation (in the form \\
   & ~~of problem \eqref{LP_relaxation}), using a LP-solver of polynomial complexity \\
6: & ~~\textbf{if} $U \leq \left\lceil {g({\mathbf{z}}_\mathcal{V}^{{\text{LP}}};{{{\mathbf{\bar z}}}_\mathcal{C}})} \right\rceil$ \textbf{then} \{\textbf{continue}\} \textbf{end if}     \hfill $\triangleright$ \textit{Pruning} \\
7: & ~~\textbf{if} ${\mathbf{z}}_\mathcal{V}^{{\text{LP}}} \in {\{ 0,1\} ^V}$ \textbf{then}    \hfill $\triangleright$ \textit{Fathoming (integer solution), given that} \\
& \hfill \scalebox{0.84}{$U > \left\lceil {g({\mathbf{z}}_\mathcal{V}^{{\text{LP}}};{{{\mathbf{\bar z}}}_\mathcal{C}})} \right\rceil  = g({\mathbf{z}}_\mathcal{V}^{{\text{LP}}};{{\mathbf{\bar z}}_\mathcal{C}}) = g({\mathbf{z}}_\mathcal{V}^{{\text{opt}}};{{\mathbf{\bar z}}_\mathcal{C}})$} \\
8: & ~~~~~$U: = g({\mathbf{z}}_\mathcal{V}^{{\text{LP}}};{{\mathbf{\bar z}}_\mathcal{C}})$, ${{\mathbf{z}}^ *}: = [{\mathbf{z}}_\mathcal{V}^{{\text{LP}}};{{{\mathbf{\bar z}}}_\mathcal{C}}]$, \textbf{continue} \\
9: & ~~\textbf{end if} \\
10: & ~~Compute a feasible solution ${\mathbf{z}}_\mathcal{V}^{{\text{CF}}}$ of the examined subproblem, \\ 
    & ~~using the CF-based greedy method (Algorithm 1)       \\
11: & ~~\textbf{if} $g({\mathbf{z}}_\mathcal{V}^{{\text{CF}}};{{\mathbf{\bar z}}_\mathcal{C}}) < U$ \textbf{then} \{$U: = g({\mathbf{z}}_\mathcal{V}^{{\text{CF}}};{{\mathbf{\bar z}}_\mathcal{C}})$, ${{\mathbf{z}}^ *}: = [{\mathbf{z}}_\mathcal{V}^{{\text{CF}}};{{{\mathbf{\bar z}}}_\mathcal{C}}]$\} \textbf{end if} \\
12: & ~~\textbf{if} $\left\lceil {g({\mathbf{z}}_\mathcal{V}^{{\text{LP}}};{{{\mathbf{\bar z}}}_\mathcal{C}})} \right\rceil  = g({\mathbf{z}}_\mathcal{V}^{{\text{CF}}};{{\mathbf{\bar z}}_\mathcal{C}})$ \textbf{then} \{\textbf{continue}\} \textbf{end if}    \hfill $\triangleright$ \textit{Fathoming} \\
13: & ~~Select a branching variable ${z_b}$ $\left( b: = \mathop {\arg \min }\limits_{v \in \mathcal{V}} \left| {z_v^{{\text{LP}}} - 0.5} \right| \right)$, and then \\
    & ~~generate two new subproblems in the form of problems \eqref{BILP_subproblem_0} and \eqref{BILP_subproblem_1} \\
14: & ~~Insert the generated subproblems at the end of the list $\mathcal{L}$ \\
15: & \textbf{end while}  \\       
\end{tabular}
\\ \hline
\end{tabular*}
\vspace{-4mm}
\end{table}

The proposed B\&B method is given in Algorithm 2, where $U$ is the best \textit{global upper bound} found so far by the algorithm $(g^* \leq U)$, and $\mathcal{L}$ is the \textit{list of active subproblems} that controls the order in which the subproblems are examined (a generated subproblem is called active if it has not been examined yet). Note that $\mathcal{L}$ is a first-in-first-out (FIFO) list; this is preferable when ``good'' upper bounds are available in order to ``prune'' the search space as early as possible. 

Furthermore, the B\&B method performs three \textit{fundamental operations}, where no further investigation is needed for the examined subproblem: 1) \textit{Infeasibility}: the examined subproblem is infeasible, 2) \textit{Pruning}: the examined subproblem cannot produce a better solution $(U \leq \left\lceil {g({\mathbf{z}}_\mathcal{V}^{{\text{LP}}};{{{\mathbf{\bar z}}}_\mathcal{C}})} \right\rceil)$, and 3) \textit{Fathoming}: an optimal solution of the examined subproblem is found; this occurs when the solution of the LP relaxation is integer $({\mathbf{z}}_\mathcal{V}^{{\text{LP}}} \in {\{ 0,1\} ^V})$, or when $\left\lceil {g({\mathbf{z}}_\mathcal{V}^{{\text{LP}}};{{{\mathbf{\bar z}}}_\mathcal{C}})} \right\rceil  = g({\mathbf{z}}_\mathcal{V}^{{\text{CF}}};{{\mathbf{\bar z}}_\mathcal{C}})$ which implies $g({\mathbf{z}}_\mathcal{V}^{{\text{opt}}};{{\mathbf{\bar z}}_\mathcal{C}}) = g({\mathbf{z}}_\mathcal{V}^{{\text{CF}}};{{\mathbf{\bar z}}_\mathcal{C}})$. Finally, Algorithm 2 produces a \textit{nonincreasing sequence of global upper bounds} $U$, and after its termination $U = g^*$ since all the generated subproblems have been examined ($\mathcal{L} = \emptyset$).

\textit{Complexity of Algorithm 2}: The complexity of each iteration is mainly restricted by the LP-solver (polynomial complexity $O( {{(T + V)^{1.5}}{V^2}} )$ \cite{Ben-Tal}) as well as Algorithm 1, so it is $O( {{(T + V)^{1.5}}{V^2} + T{V^2}} ) = O( {{(T + V)^{1.5}}{V^2}} ) = O( {{(T + K)^{1.5}}{K^2}} )$. Furthermore, in each iteration we examine one subproblem, while we generate at most two new subproblems by fixing one of the free variables. Therefore, the number of iterations/subproblems is $\leq \sum\limits_{j = 0}^K {{2^j}}  = 2^{K + 1} - 1 = O(2^K)$. Overall, the worst-case complexity of Algorithm 2 is $O( {{2^K}{(T + K)^{1.5}}{K^2}} )$, i.e., exponential in the size of the input. Although the original BILP problem \eqref{original_BILP_problem} is probably intractable in the worst case (due to its NP-hardness), the most difficult problem instances may rarely occur in practice (because of their special structure), so the \textit{average-case complexity} may be a more appropriate measure of an algorithm's efficiency. Specifically, assuming a probability distribution over problem instances, the average-case complexity of Algorithm 2 is $O( {M{(T + K)^{1.5}}{K^2}} )$, where $M$ is the mean/average number of iterations. Observe that if $M = \operatorname{poly}(T,K)$, where $\operatorname{poly}(T,K)$ is some polynomial in $T$ and $K$, then Algorithm 2 will have polynomial-time complexity in the average case.\footnote{Note that an \textit{exhaustive-enumeration algorithm}, despite its global optimality, requires $T\sum\limits_{j = 1}^K {\binom{K}{j}} = T( {2^K} - 1 ) = \Theta ({2^K}T)$ comparisons in all cases, thus having exponential complexity in both the worst and the average case.}  Nevertheless, the average-case complexity of the B\&B method is very challenging to study theoretically, so we resort to a numerical analysis in Section V.

\section{Numerical Results and Discussion}
In this section, the performance of the designed B\&B algorithm is evaluated through a series of problem instances. More specifically, the following simulation parameters have been considered: $K \in \{ 10,15,20,25,30\}$, $T = 12$, $P_t^{{\text{avl,req}}} = 99.9\%$, $\forall t \in \mathcal{T}$, and 200 independent scenarios (for each value of $K$) with the outage probabilities $\{p_{k,t}^{{\text{out}}}\}_{ k \in \mathcal{K}, \, t \in \mathcal{T} }$ being uniformly distributed in the interval $[0.1,1]$.

\begin{table}[!t]
\begin{threeparttable}[b]
\caption{Performance Comparison with Existing Methods: Average \# of Selected GSs \& Percentage of Problems Optimally Solved\,\textsuperscript{\normalfont a}}
\centering
\renewcommand{\arraystretch}{1.35}
\begin{tabular}{|c|c|c|c|c|}
\hline
\diagbox[width=1.54cm, height=0.95cm]{$K$}{\scalebox{0.6}{METHOD}} & ESA \cite{Lyras_GEO} & GHA\tnote{b} \ \cite{Lyras_GEO} & CHA\tnote{b} \ \cite{Lyras_GEO} & Algorithm 2 \\  \hline
10 & 9.84 & 9.89 (96\%) & 9.87 (97\%) & 9.84 (100\%) \\ \hline
15 & 11.36 & 11.89 (55\%) & 11.70 (68\%) & 11.36 (100\%) \\ \hline
20 & 10.33 & 11.15 (33\%) & 10.74 (60\%) & 10.33 (100\%) \\ \hline
25 & 9.62 & 10.62 (17\%) & 10.00 (63\%) & 9.62 (100\%) \\ \hline
30 & 9.23 & 10.14 (24\%) & 9.65 (58\%) & 9.23 (100\%) \\ \hline
\end{tabular}
\begin{tablenotes}
\item[a] This percentage is calculated using the global minimum obtained from the exhaustive-search algorithm (ESA) given in \cite{Lyras_GEO}.
\item[b] GHA and CHA select up to 3 and up to 2 redundant GSs, respectively.
\end{tablenotes}
\end{threeparttable}
\vspace{-2mm}
\end{table}

\begin{table}[!t]
\caption{Iterations Required by Algorithm 2}
\centering
\renewcommand{\arraystretch}{1.35}
\begin{tabular}{|c|c|c|c|}
\hline
$K$ & \scalebox{1.0}{\makecell{Total \# of \\ iterations \\ $[$mean (standard \\ deviation)$]$}} & \scalebox{1.0}{\makecell{\# of iterations until a \\ global minimum is found \\ for the 1\textsuperscript{st} time $[$mean \\  (standard deviation)$]$}} & \scalebox{1.0}{\makecell{Upper bound on \\ the total \# of \\ iterations \\ $[= {2^{K + 1}} - 1]$}} \\ \hline
10 & 1.93 (2.37) & 0.22 (0.71) & $ > 2 \times {10^3}$ \\ \hline
15 & 14.23 (15.61) & 6.04 (10.30) & $ > 6 \times {10^4}$ \\ \hline
20 & 41.37 (57.01) & 15.86 (38.91) & $ > 2 \times {10^6}$ \\ \hline
25 & 87.74 (117.14) & 27.52 (75.55) & $ > 6 \times {10^7}$ \\ \hline
30 & 117.90 (204.85) & 28.42 (121.32) & $ > 2 \times {10^9}$ \\ \hline
\end{tabular}
\vspace{-4mm}
\end{table}

Firstly, we compare Algorithm 2 with state-of-the-art methods. As shown in Table I, GHA exhibits the lowest performance, while Algorithm 2 achieves exactly the same performance with ESA and significantly outperforms GHA and CHA. Moreover, for $K \in \{ 15,20,25,30\}$, GHA and CHA attain a globally optimal solution in less than 70\% of cases.\footnote{Although the worst-case complexity of both GHA and CHA is $\Theta ({T{K^2}})$, these heuristic methods do not provide any performance guarantee.} On the other hand, Algorithm 2 finds the global optimum in all cases, since it is theoretically guaranteed to do so.

Furthermore, we examine the complexity of Algorithm 2 in terms of the required iterations (recall that each iteration has polynomial-time complexity). According to Table II, the B\&B method requires extremely few iterations on average (with small standard deviation) compared to the upper bound $2^{K + 1} - 1$. Thus, Algorithm 2 has low average-case complexity.

\begin{figure}[!t]
\centering
\includegraphics[width=2.82in]{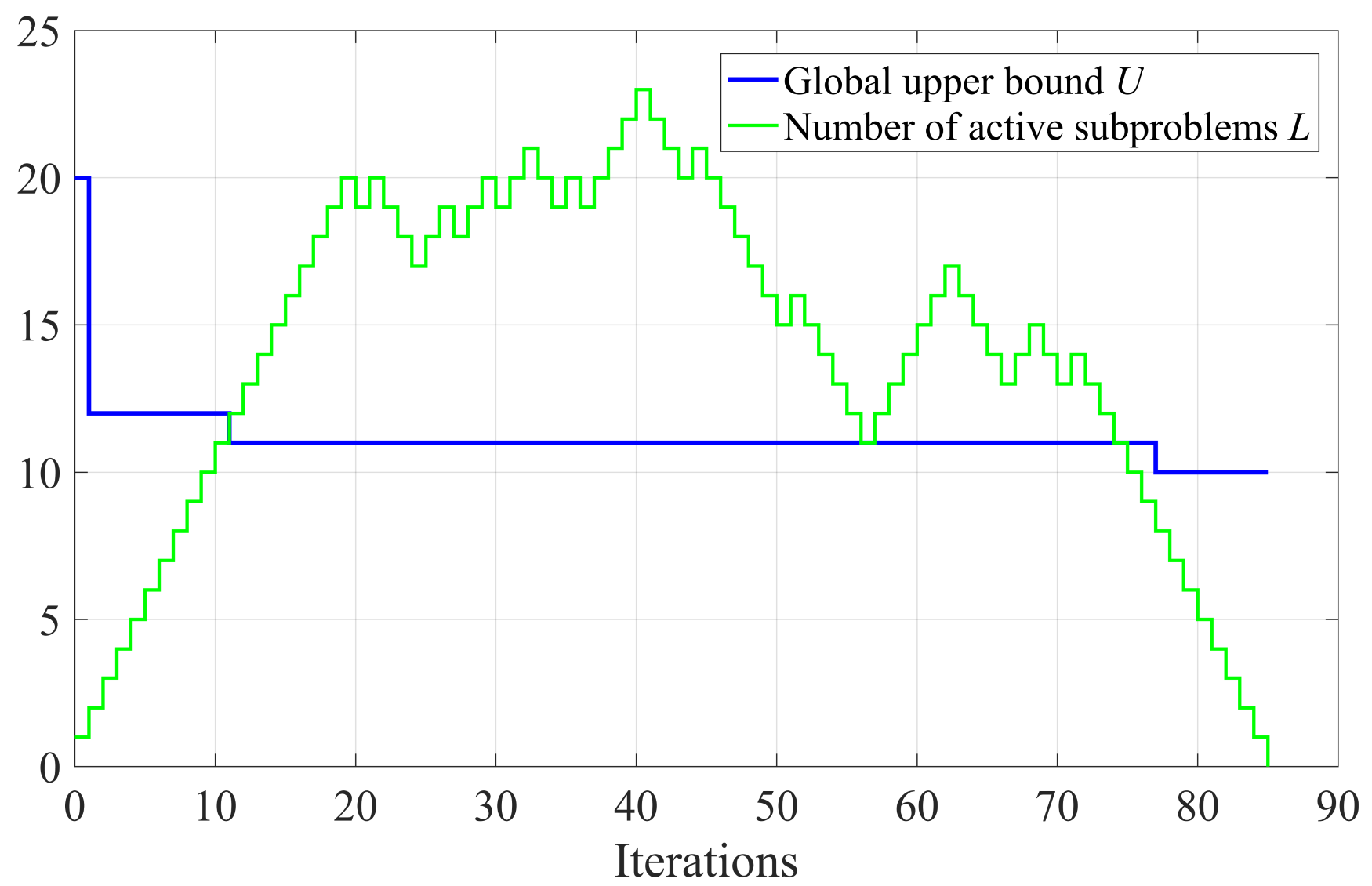} 
\caption{Progress of Algorithm 2 for a simulation scenario with $K = 20$. Global minimum $=$ 10, total number of iterations $=$ 85, and number of iterations until a global minimum is found for the 1\textsuperscript{st} time $=$ 77.}
\label{Fig1}
\vspace{-4mm}
\end{figure}

Finally, Fig. 1 illustrates the progress of the B\&B method for a specific problem. In particular, we can observe: 1) the nonincreasing sequence of global upper bounds $U$, and 2) that the number of active subproblems $\left| \mathcal{L} \right| = L$ is equal to 1 at the beginning of the algorithm, and becomes 0 in the last iteration. 

\section{Conclusion}
In this article, we have studied the optimal selection of GSs in satellite systems with site diversity. Furthermore, we have developed a global optimization algorithm, which can provide significant cost savings for the network operator. Finally, according to the numerical results, the proposed B\&B method exhibits low average-case complexity, while achieving much higher performance than existing algorithms.


\end{document}